\documentclass[]{aastex63}

\published{August 2023 in PSJ}
\usepackage{courier}
\usepackage{stix, isomath,amsmath}
\graphicspath{./figures}

\begin{document}

\title{Thermal properties of the leading hemisphere of Callisto inferred from ALMA observations}

\correspondingauthor{Maria Camarca}
\email{mcamarca@caltech.edu}
\author[0000-0003-3887-4080]{Maria Camarca}
\affiliation{Division of Geological and Planetary Sciences, California Institute of Technology 1200 E California Blvd M/C 150-21 Pasadena, CA 91125, USA}
\author[0000-0002-9068-3428]{Katherine de Kleer}
\affiliation{Division of Geological and Planetary Sciences, California Institute of Technology 1200 E California Blvd M/C 150-21 Pasadena, CA 91125, USA}
\author[0000-0002-5344-820X]{Bryan Butler}
\affiliation{National Radio Astronomy Observatory, Socorro, NM 87801, USA}
\author[0000-0001-8379-1909]{Alex B. Akins}
\affiliation{Jet Propulsion Laboratory, California Institute of Technology, Pasadena, CA 91011, USA}
\author[0000-0002-8178-1042]{Alexander Thelen}
\affiliation{Division of Geological and Planetary Sciences, California Institute of Technology 1200 E California Blvd M/C 150-21 Pasadena, CA 91125, USA}
\author[0000-0002-4278-3168]{Imke de Pater}
\affiliation{Department of Astronomy, UC Berkeley, Berkeley, CA 94720, USA}
\author[0000-0003-0685-3621]{Mark A. Gurwell}
\affiliation{Center for Astrophysics | Harvard \& Smithsonian, 60 Garden Street, Cambridge, MA 02138, USA}
\author[0000-0002-9820-1032]{Arielle Moullet}
\affiliation{National Radio Astronomy Observatory, Charlottesville, VA 22903, USA}

\begin{abstract}
We present a thermal observation of Callisto’s leading hemisphere obtained using the Atacama Large Millimeter/submillimeter Array (ALMA) at 0.87 mm (343 GHz). The angular resolution achieved for this observation was $\sim$$0.16^{\prime\prime}$, which for Callisto at the time of this observation ($D\sim 1.05^{\prime\prime}$) was equivalent to $\sim$6 elements across the surface. Our disk-integrated brightness temperature of 116 $\pm$ 5 K (8.03 $\pm$ 0.40 Jy) is consistent with prior disk-integrated observations. Global surface properties were derived from the observation using a thermophysical model \citep{de_kleer_ganymedes_2021} constrained by spacecraft data. We find that models parameterized by two thermal inertia components more accurately fit the data than single thermal inertia models. Our best-fit global parameters adopt a lower thermal inertia of 15-50 $\text{J}\:\text{m}^{-2}\:\text{K}^{-1}\:\text{s}^{-1/2}$ and a higher thermal inertia component of 1200-2000 $\text{J}\:\text{m}^{-2}\:\text{K}^{-1}\:\text{s}^{-1/2}$, with retrieved millimeter emissivities of 0.89-0.91. We identify several thermally anomalous regions, including spots $\sim$3 K colder than model predictions co-located with the Valhalla impact basin and a complex of craters in the southern hemisphere; this indicates the presence of materials possessing either a higher thermal inertia or a lower emissivity. A warm region confined to the mid-latitudes in these leading hemisphere data may be indicative of regolith property changes due to exogenic sculpting.
\end{abstract}



\section{Introduction} 

The Galilean moons of Jupiter – Io, Europa, Ganymede, and Callisto – form a continuum of geologic activity for which the inactive endmember is Callisto.  A satellite whose surface is dominated almost exclusively by impact craters, Callisto retains one of the oldest surfaces ($\sim$4.5 Gyr) in the Solar System. Although it is the third largest planetary moon and may harbor a subsurface ocean \citep{zimmer_subsurface_2000}, Callisto bears little to no evidence for any tectonics, volcanism, or other manifestations of surface-interior interactions \citep{moore_callisto_2004}. By contrast, geologic activity is readily observed on the other Galilean satellites. For instance: Io is volcanically active in the present day and thermal emission from its colorful eruptions is routinely observed using ground-based telescopes \citep{de_kleer_ios_2019} and spacecraft \citep{mura_infrared_2020,davies_heartbeat_2006}; Europa has a young, water-rich surface accompanied by evidence for surface-interior interactions \citep{pappalardo_europas_2009}; and Ganymede is encircled by bright, ribbon-like, tectonically-deformed regions suggestive of major past activity \citep{bagenal_ganymede_2004}. Therefore, as the only Galilean moon with a surface dominated by exogenic sculpting, Callisto stands out as one of the best records of long-term impact modification the Solar System has to offer.

As a consequence of its inactivity, Callisto's surface at the global scale comprises just a few geologic units. These units include cratered plains, light plains, and smooth/dark plains, as well as units linked to multi-ring basins or other impacts \citep{moore_callisto_2004}. In particular, Callisto's multi-ring impact basins are the largest in the Solar System. Valhalla, the largest of these basins, has outer rings extending to a diameter of $\sim$3800 km and occupies $\sim$16\% of Callisto's surface area. Structures like Valhalla are possible relics of a time in Callisto's history where the surface had a higher thermal gradient and was more ductile \citep{schenk_geology_1995} and represent regions of interest to look for exposed subsurface materials. While the multi-ring basins are the most obvious large features on Callisto, the dominant landform is the heavily cratered plains. These swaths of impact-scarred terrain are often blanketed by a dark material for which the exact composition and origin remains uncertain, with endogenic explanations often focusing on surface erosion/sublimation processes \citep{moore_callisto_2004} and exogenic theories usually invoking dust infall from the irregular Jovian satellites \citep{bottke_black_2013}. Although Callisto retains the largest share of dark material (its albedo is 0.2, the lowest of the Galileans, \citealt{moore_callisto_2004}), Europa and Ganymede also have similarly dark products on their surfaces \citep{bagenal_geology_2004,bagenal_ganymede_2004} . Altogether, these features contribute to the fact that Callisto’s quiescent surface remains poorly understood.

For many Solar System objects, unknowns pertaining to unique geologic features or the balance of exo- and endogenic processes on their surfaces have been investigated by inferring material properties from thermal observations. At wavelengths in the infrared and (sub)millimeter, radiation originates from the surface/near-subsurface and can be leveraged to infer properties such as emissivity, thermal inertia ($\Gamma = \text{J}\:\text{ m}^{-2}\:\text{ K}^{-1}\:\text{ s}^{-1/2}$; henceforth thermal inertia will be given without units but always uses these SI units), porosity, etc. The variation of such properties across a surface encodes information as to how that surface evolved and provides clues regarding the nature of the external modification environment. For example, in the Saturnian system, thermal infrared Cassini Composite Infrared Spectrometer (CIRS) measurements of the small icy moons Tethys \citep{howett_pacman_2012, howett_maps_2019} and Mimas \citep{howett_high-amplitude_2011} revealed both satellites have a high thermal inertia feature centered at mid-latitudes on their leading hemispheres, demonstrating the Saturnian particle environment is capable of altering surface texture \citep{howett_high-amplitude_2011,schaible_high_2017,schenk_plasma_2011}. At the Moon, measurements across the $\sim$13-400 µm regime from the Diviner Lunar Radiometer onboard the Lunar Reconnaissance Orbiter showed that the Moon's regolith is uniform at global scales, suggesting processes like impact gardening work quickly to homogenize the upper 10 cm; additionally, maps of local thermal inertia variations helped assign plausible impact ejecta deposits to parent craters \citep{hayne_global_2017}. In the Jovian system, recent work with the Atacama Large Millimeter/submillimeter Array (ALMA) has proved useful for characterizing the surface of Io \citep{de_pater_alma_2020}, confirming the surficial origin of apparent nighttime hot spots on Europa \citep{trumbo_alma_2017,trumbo_alma_2018}, and identifying plausible exogenic thermal trends on Ganymede \citep{de_kleer_ganymedes_2021}. With all of these studies, surface trends were identified with hemisphere-scale or global-scale coverage of the planetary body. In particular, some (sub)millimeter/centimeter observations may probe below the strongest time-of-day temperature variations (i.e., below the diurnal thermal skin depth) and below the most heavily processed upper surface layers. 

At present, the thermal properties of Callisto's surface are poorly constrained because there are only a few disk-resolved studies at thermal wavelengths. In the infrared, the Voyager Infrared Interferometer Spectrometer and Radiometer (IRIS) instrument returned a featureless spectrum of Callisto, offering little information regarding composition \citep{spencer_surfaces_1987}. However, the eclipse cooling curves showed that Callisto’s surface is not thermally uniform, as the best-fit model adopted a two-component thermal inertia surface with one higher component of $\Gamma$ = 300 and a lower component of $\Gamma$ = 15 \citep{spencer_surfaces_1987}. Moreover, the IRIS spectra exhibited a trend of increasing brightness temperature with decreasing wavelength, suggesting unresolved temperature contrasts on the surface; this trend also varied with solar incidence angle, hinting that Callisto's surface may be less smooth than Ganymede's, which did not exhibit this solar incidence angle dependence \citep{spencer_surfaces_1987}. The  thermal IR temperature of Callisto’s surface was measured to be $\sim$158 K at 20 µm \citep{morrison_temperatures_1972,moore_callisto_2004} and $\sim$140-150 K at $\sim$10 µm \citep{de_pater_sofia_2021}, while the H\textsubscript{2}O ice surface temperature derived using H\textsubscript{2}O spectral features is somewhat lower at $\sim$115 K \citep{grundy_near-infrared_1999}. Past disk-integrated radio thermal wavelength measurements of Callisto's brightness temperature show a decrease from 135 K \citep{de_pater_planetary_1989} in the sub-mm regime to about $\sim$90-100 K at cm wavelengths \citep{pauliny-toth_brightness_1974,muhleman_precise_1986,butler_alma_2012,berge_callisto_1975, de_pater_vla_1984}. Although other disk-integrated results for Callisto fill in the gaps between these wavelengths, there is a scarcity of spatially resolved datasets.

In this work, we present the first high-resolution thermal observation of Callisto using ALMA by mapping the leading hemisphere at 0.87 mm (343 GHz). Section \ref{Methods} describes the observations, data analysis, and flux calibration. Section \ref{Thermal Model} details the thermophysical model used to interpret the data and describes how surface properties are derived. Section \ref{Results & Discussion} presents the final, calibrated leading hemisphere image of Callisto as obtained with ALMA and includes the results and interpretation of the thermophysical modeling analysis. Section \ref{Conclusion} summarizes the conclusions of this work.
\section{Methods} \label{Methods}
\subsection{Observations}
Our observations of Callisto were acquired using ALMA. Located on the Chajnantor plateau in the Atacama desert, Chile, the main ALMA array is composed of 54 12-meter radio antennas that are linked via a correlator to function as an interferometer. As an interferometer, ALMA can achieve higher spatial resolution on planetary objects than is possible using single-dish facilities at the same wavelength. The spatial resolution ALMA can achieve, combined with its frequency coverage, make this telescope facility ideal for conducting thermal studies of the Galilean satellites from Earth.
   
We observed the leading hemisphere of Callisto using ALMA on 2016 November 01. The data were taken using the Band 7 receiver at a central frequency of 343 GHz (0.87 mm) with four spectral windows collectively spanning $\sim$8 GHz and an on-source integration time of 121 s. During observations, the array was in configuration C-5 with 39 12-m antennae. The baselines ranged from 18.6 m to 1.1 km, affording an angular resolution of $\sim$$0.16^{\prime\prime}$. The angular diameter of Callisto during data acquisition was $\sim$$1.05^{\prime\prime}$, so $\sim$6 resolution elements across the disk were achieved, equivalent to a spatial footprint of $\sim$790 km at the satellite's distance. At the time of the observation, the sub-observer longitude was 50$^{\circ}$ W. Calibration for array pointing, flux density, bandpass response, and phase were obtained via observations of quasars J1256-0547 and J1232-0224.

\subsection{Data Analysis}
After observations, the data were provided as a pipeline-calibrated measurement set containing the visibilities, which are the fundamental quantities measured by an interferometer and are equal to the Fourier transform of the sky brightness temperature distribution at discrete spatial frequencies \citep{muders_alma_2014}. For a review of the interferometric observation and imaging of Solar System objects, see \cite{taylor_solar_1999}. Because continuum observations of the Galilean satellites typically have very high signal-to-noise ratios (SNR) and easily modeled shapes (disks), we improved the SNR of the final pipeline-generated images by applying an iterative self-calibration procedure (for an in-depth explanation of self-calibration, see \cite{brogan_advanced_2018}). To conduct this processing, we used the Common Astronomy Software Application (CASA) package \citep{mcmullin_casa_2007}. As ALMA observations at 0.87 mm (343 GHz) probe continuum emission from the subsurface, we integrated the signal over the entire 8 GHz of bandwidth using multi-frequency synthesis \citep{sault_multi-frequency_1994}. Our phase-only self-calibration using CASA followed as such: for the first round of imaging, a Lambertian disk scaled to the size and brightness temperature of Callisto was used as a startmodel for the \texttt{tclean} task \citep{rau_multi-scale_2011}; no additional clean components were added. Next, the complex antenna-based gains were calculated using the CASA \texttt{gaincal} task on a interval spanning the full integration time (i.e., $>$ 121 s) and then applied using \texttt{applycal}. Then, the corrected visibilities were imaged again adopting a shallow clean, using the previous \texttt{tclean} output image as the startmodel. This cycle of cleaning and computing gains continued down to an interval of 2 s, at which the image SNR ceased to substantially improve. A dynamic range of $\sim$580 (peak flux/rms in units of Jy/beam) was obtained for the final image. For \texttt{tclean} parameters, we adopted a Briggs weighting scheme with a robust parameter of 0.0, and used the "Clark" deconvolver \citep{clark_efficient_1980}. A primary beam correction was applied to the self-calibrated image product. The dimensions of the final clean beam were $0.16^{\prime\prime} \times 0.19^{\prime\prime}$. The rms (root-mean-square) noise of the final image (measured from a non-source region of the non-primary beam corrected product) was 0.585 mJy/beam, or about 0.2 K.

\begin{figure}
\centering
\includegraphics[width=0.48\textwidth]{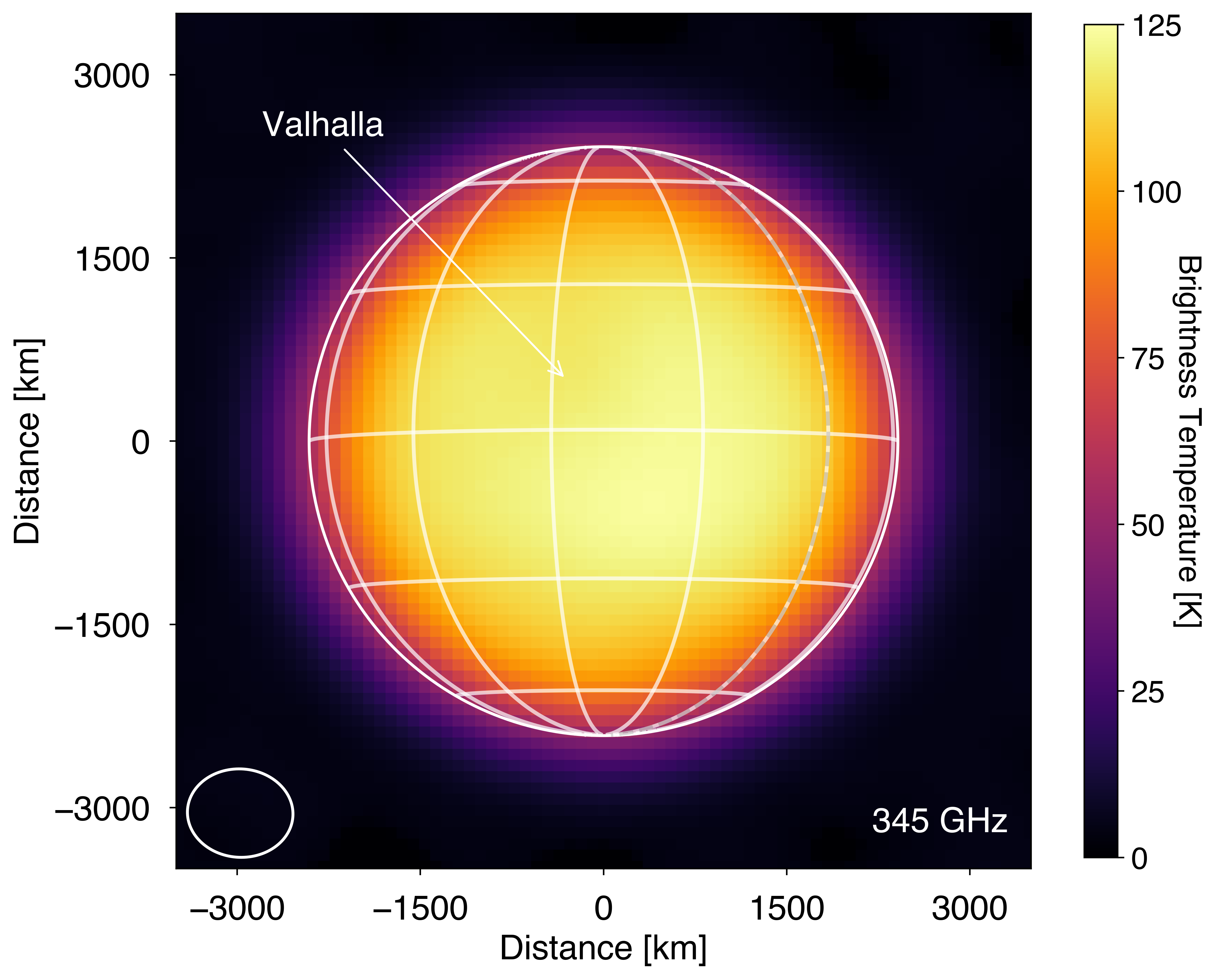}
\hfill
\includegraphics[width=0.48\textwidth]{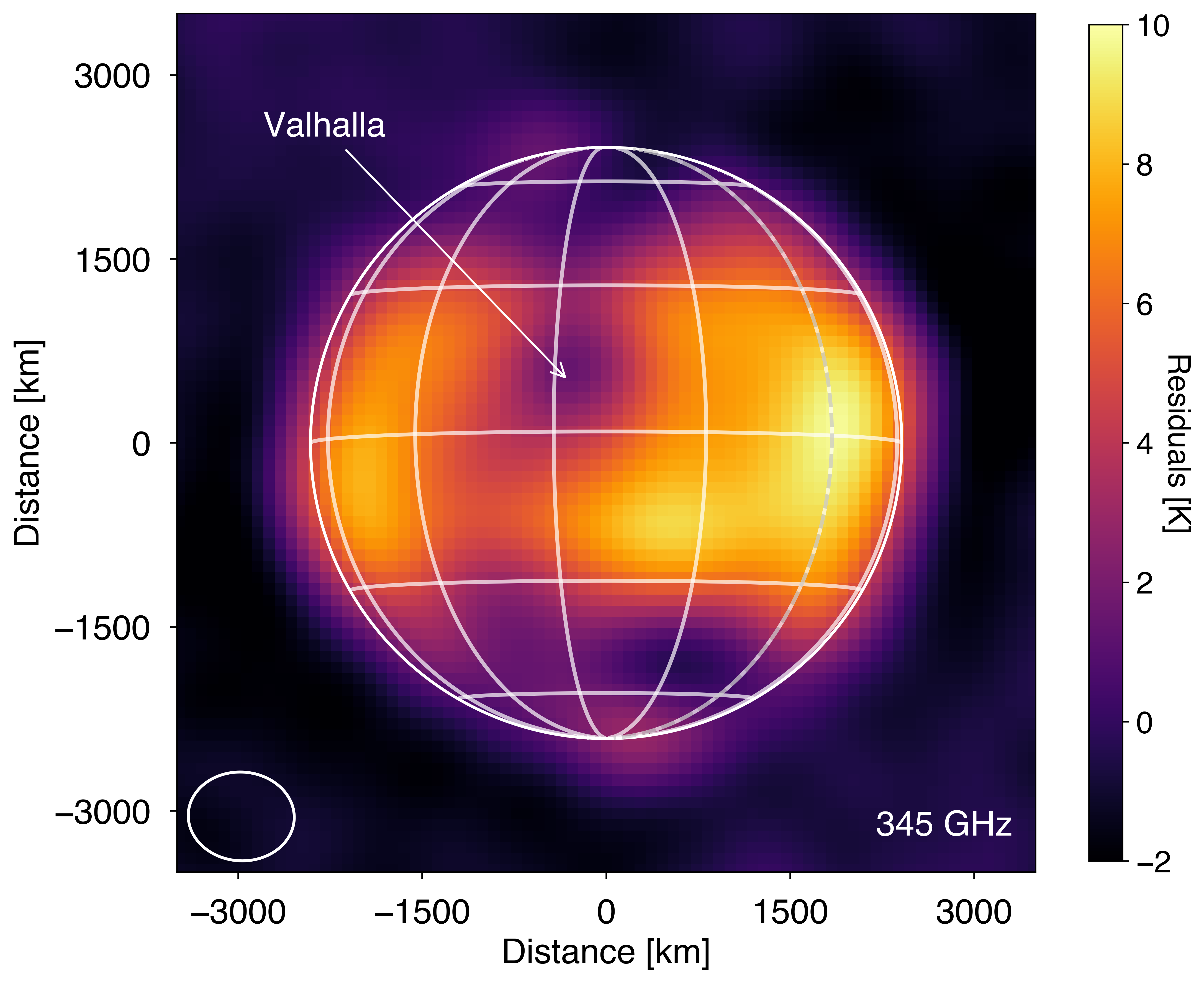}
\caption{(a) Calibrated ALMA Band 7 (0.87 mm/343 GHz) image of Callisto's leading hemisphere. (b) Result of subtracting a sample Lambertian disk from (a) to highlight difference in $T_b$ across the disk. In both images, the ellipse in the lower left corner represents the FWHM (full-width at half-maximum) of the synthesized ALMA beam. The latitude/longitude grid is spaced at 30$^{\circ}$ increments, and the 0$^{\circ}$ longitude line is marked with dashes. Callisto's north pole is aligned with the image axis. Notably, the Valhalla impact crater appears slightly colder than other disk regions viewed at similar incident and emission angles.}
\label{fig:data}
\end{figure}

After self-calibrating the data, we derived the disk-integrated flux density directly from the visibilities. We report disk-integrated flux densities from these fits rather than from the images because the visibilities are the more direct data product. Reporting the disk-integrated flux is useful for placing our results in context with previous radio observations of Callisto that are largely spatially unresolved. The visibilities were fit with Bessel functions using the CASA task \texttt{uvmodelfit} assuming a uniform disk model, fitting to all spectral windows, and by excluding baselines $>$200 m for an optimal fit, as the longer baselines are sensitive to the variations across the disk, rather than just the disk-integrated flux. The final disk-integrated brightness temperature ($T_b$) was calculated by inverting the following equation \footnote{This equation has a typo in \cite{de_kleer_ganymedes_2021}, but is correct in the latest arXiv version of that paper (\url{https://arxiv.org/pdf/2101.04211.pdf}). The '206265', which is the number of arcseconds in a radian, was not squared.}:

\begin{equation} F{_\nu} = 10^{26}\frac{\pi R^2_C}{206265^{2}}\frac{2h\nu^3}{c^2}\times [\frac{1}{e^{h\nu/k_bT_b}-1} -\frac{1}{e^{h\nu/k_bT_{cmb}}-1} ] \end{equation}
where $F{_\nu}$ is the disk-integrated flux density in Janskys, $R_C$ is the radius of Callisto in arcseconds and, in SI units, $h$ is Planck's constant, $c$ is the speed of light, $\nu$ is the observation frequency, $k_b$ is the Boltzmann constant, and $T_{cmb}$ is the cosmic microwave background (2.7 K). The uncertainties in the final $T_b$ incorporate flux density scale calibration errors, which was parameterized by a 5\% error of the total flux density, and visibility fitting errors on the order of 0.4 mJy. Our final calibrated image is presented in Fig.~\ref{fig:data}.  

\subsubsection{ALMA Flux Density Calibration Check}
The ALMA pipeline applies a flux density calibration to the target data during initial processing. Because the ALMA calibrator sources are quasars (which are variable) and the ALMA pipeline occasionally does not use the most complete calibrator information, manual flux calibration correction is sometimes required  (e.g., \citealt{trumbo_alma_2018,de_kleer_ganymedes_2021}). We checked whether such a correction was required for these data by comparing the pipeline flux model to the ALMA Calibrator Catalogue\footnote{\url{https://almascience.nrao.edu/sc/}} of all observations of our calibration source in the period surrounding our observations.  The quasar used for flux calibration was J1256-0547, which the calibrator catalogue recorded observations of at Bands 3 and 7. A check of flux density measurements in these bands between 2016-10-27 and 2016-11-01 (i.e., before and after our data were obtained) indicated it is not necessary to correct the pipeline calibration of these data. 

\section{Thermophysical Model}\label{Thermal Model}
The final product of calibrating and imaging the ALMA Callisto visibilities is a map of brightness temperature across the satellite's disk (Fig.~\ref{fig:data}). In order to interpret this observed $T_b$ map, we fit the data with a thermophysical model that treats the transport of heat by conduction and radiation through Callisto's surface. For each latitude and longitude on the satellite, the model constructs a temperature profile that evolves with time \emph{t} and depth \emph{z} into the surface. Practically, the thermophysical model numerically solves the 1D heat diffusion equation, which is given by: 

\begin{equation} \rho c{_p} \frac{\partial T}{\partial t} = \frac{\partial}{\partial z}(k \frac{\partial T}{\partial z}) \end{equation} where $\rho$ is the density, $c{_p}$ is the heat capacity, $T$ is the temperature, and $k$ is the thermal conductivity. For this differential equation, the boundary conditions adopted no heat flow at depth and solar insolation at the distance of Callisto based on spatially varying albedo, as described in Section \ref{albedo-map}. The model was evolved over $\sim$15 Callisto days (1 Callisto day = 16.69 Earth days) in time steps of at least 1/500 of a day to allow for equilibration, and was run up to several thermal skin depths below the surface. The key tunable parameters in the model are the thermal inertia and emissivity. To find the best-fit model for our Band 7 Callisto image, we performed model runs over a wide range of thermal inertias, ranging from $\Gamma$ = 15 to $\Gamma$ = 2000, corresponding to values appropriate for very unconsolidated material and water ice, respectively \citep{ferrari_low_2016}. For each model, we integrated the emission at depth along the line of sight to account for subsurface radiation that is probed at ALMA thermal wavelengths. For full details of the model, including a full description of the radiative transfer, dielectric properties, and other key processes and equations, see Section 3 of \cite{de_kleer_ganymedes_2021}. The final output of the model is an image of integrated thermal emission from Callisto's (sub)surface at the viewing geometry of the observations, convolved to the resolution of the ALMA data. The model image is then compared directly to the observations to determine the best-fit thermal inertia and emissivity.
\subsection{Two-Component Mixtures}
Although some models generated using a single thermal inertia provided a better fit for either Callisto's limb or center-of-disk, no single model could fit both parts, regardless of treatment of the surface emission incidence angle dependence (e.g. by using smooth or rough surface implementations). We further considered two-component surface models, which have previously been employed to fit observations of the Galilean satellites, and Callisto specifically. For example, \cite{spencer_surfaces_1987} reported that Voyager infrared cooling curves of Callisto that displayed an initial, fast cooling followed by a slower cooling rate could be modeled by a surface with a high thermal inertia $\Gamma\sim$ 300 overlain by a thin, low thermal inertia layer of $\Gamma\sim$ 15. Here, we explored how the systematics in the single model residuals (e.g., rings on the limb in the residuals) could be addressed by a two-component approach. 

\begin{figure}
\centering
\includegraphics[scale=0.24]{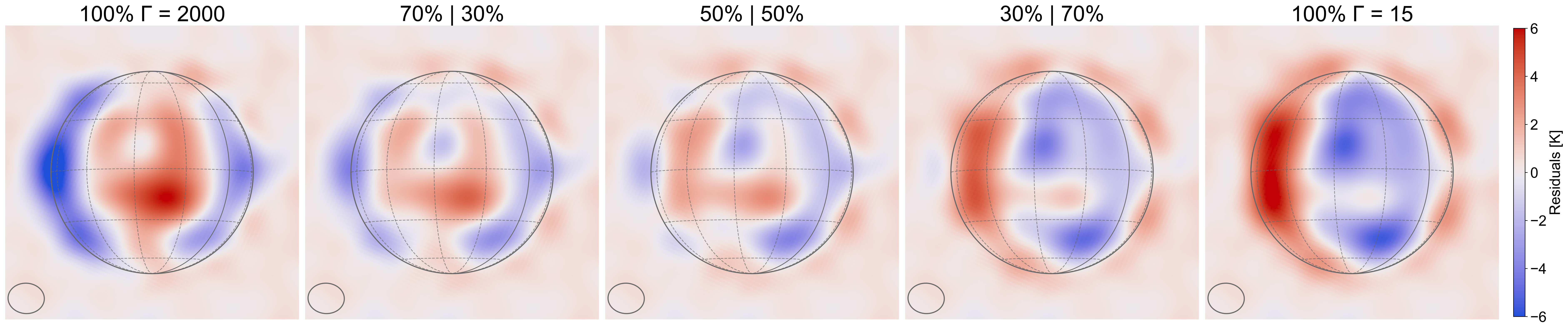}
\caption{Examples of a linear mixture of two models of different thermal inertias ($\Gamma$, $\text{J}\:\text{m}^{-2}\:\text{K}^{-1}\:\text{s}^{-1/2}$), one using $\Gamma$ = 2000 and another using $\Gamma$ = 15. Positive (red) and negative (blue) residuals correspond to surface regions in the data that are warmer and cooler than predicted by the model, respectively. The figure demonstrates that very low thermal inertia models predict too little limb emission relative to disk center, while very high thermal inertia models provide a poor fit for the opposite reason; a mixture of the two models provides a better fit without systematic center-to-limb trends in the residuals.}
\label{fig:model-comparison}
\end{figure}

To create each two-component model, we conducted the following: first, two sets of unconvolved models were generated; the first set adopted thermal inertias ranging from $\Gamma_{lower}$ = 15-400 and the second set used $\Gamma_{higher}$ = 500-2000. Each member of the low thermal inertia set was linearly combined with each member of the high thermal inertia set, forming two-component pairs. For each pair of $\Gamma_{lower}$ and $\Gamma_{higher}$ models, we created 11 linear mixtures that ranged from 0$\%$ $\Gamma_{lower}$ + 100$\%$ $\Gamma_{higher}$, to 100$\%$ $\Gamma_{lower}$ + 0$\%$ $\Gamma_{higher}$; the models were mixed in increments of 10$\%$ (see Fig.~\ref{fig:model-comparison} for an example mixture). The spatial alignment of each model pair was checked for proper overlap. After the components were mixed, the two-$\Gamma$ model was convolved with the synthesized ALMA beam and subtracted from the data. The emissivity for each mixed model was obtained by performing a least-squares analysis, with the off-disk flux masked out for the calculation. Goodness of fit for the thermophysical models was determined using the cost function, which takes into consideration the number of resolution elements (i.e., ALMA beams). Details for applying the cost function to ALMA data are described in Section 3.4 of \cite{de_kleer_surface_2021}; we note that the cost function is formally limited to data sets with independent data points.

\subsection{Albedo Map}\label{albedo-map}
\begin{figure}
\centering
\includegraphics[scale=0.5]{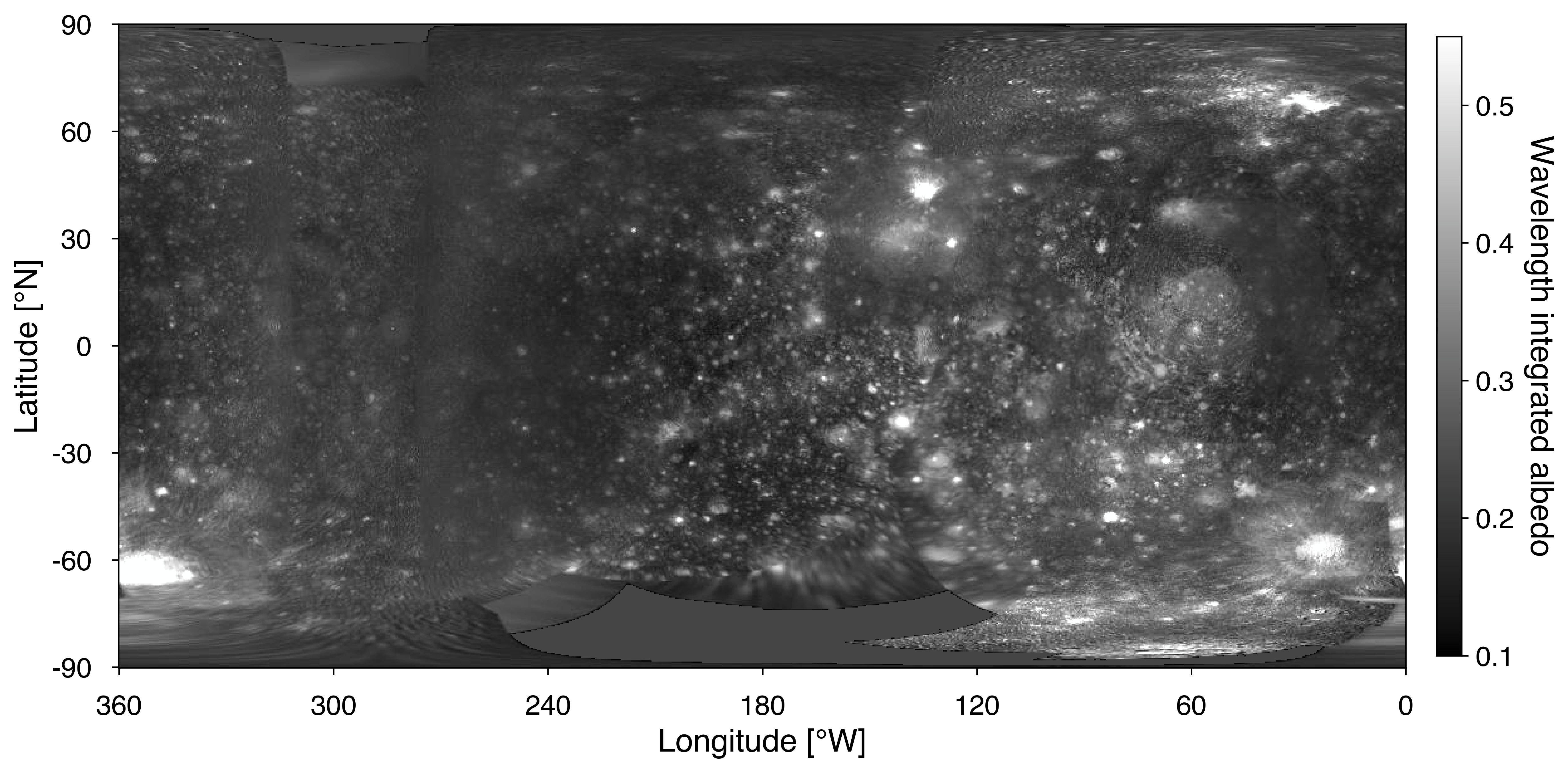}
\caption{A high-resolution version of the wavelength integrated albedo map of Callisto based on Voyager and Galileo data. The creation of this map is described in \ref{albedo-map}.}
\label{fig:albedo-map}
\end{figure}

Our model does not treat the albedo of Callisto as a free parameter, rather it ingests a bolometric (bond) albedo map derived from spacecraft data, following \cite{trumbo_alma_2017,trumbo_alma_2018} and \cite{de_kleer_ganymedes_2021}. For this map, we combined information from: (1) a high-resolution ($\sim$1 km) USGS greyscale map of Callisto’s surface\footnote{\url{https://astrogeology.usgs.gov/search/map/Callisto/Voyager-Galileo/Callisto_Voyager_GalileoSSI_global_mosaic_1km}}, with (2) spectral albedos measured at 0.35, 0.41, 0.48, and 0.59 µm obtained at 20 locations by Voyager \citep{johnson_global_1983}. First, the coordinates of the 20 spectral albedo measurements were used to identify 20 corresponding locations on the USGS map. For each of these locations on the USGS map, an average greyscale value was derived from a ~40 $\times$ 40 km box matched to the resolution of Voyager. Next, because the 0.35 - 0.59 µm range covers only $\sim$80\% of the solar spectrum, we extended the spectral coverage by scaling ~0.6 – 2.5 µm disk-integrated reflections to our data \citep{clark_galilean_1980}. Then, we computed a weighted, wavelength-integrated albedo for each of the 20 points, derived a linear function relating greyscale value to wavelength-integrated albedo, and then mapped this function to the entire USGS map (Fig.~\ref{fig:albedo-map}). The wavelength-integrated map was then multiplied by a phase integral of 0.51 \citep{buratti_ganymede_1991} to convert it to a bolometric albedo map. We replaced poorly sampled regions near the poles in the greyscale map with a global average; the exact values of the albedo map at the poles has negligible impact on the thermal model fits since the viewing geometry of the data is nearly equatorial.

\section{Results \& Discussion}\label{Results & Discussion}
We present surface properties derived from an ALMA 0.87 mm (343 GHz) observation of Callisto’s leading hemisphere, as well as brightness temperature residual maps created with best-fit thermophysical models. In Section \ref{integrated-brightness}, we present an integrated $T_b$ measurement, and place our result in context with past, unresolved thermal wavelength observations. In Section \ref{single-component-results}, we describe the single thermal inertia model results. In Section \ref{two-component-results}, we discuss the results from modeling the data with two thermal inertia components. Lastly, in Section \ref{local-residuals} we detail the connections between regional temperature signatures and the local geology.

\subsection{Disk-Integrated Brightness Temperature}\label{integrated-brightness}
\begin{figure}
\centering
\includegraphics[scale=0.43]{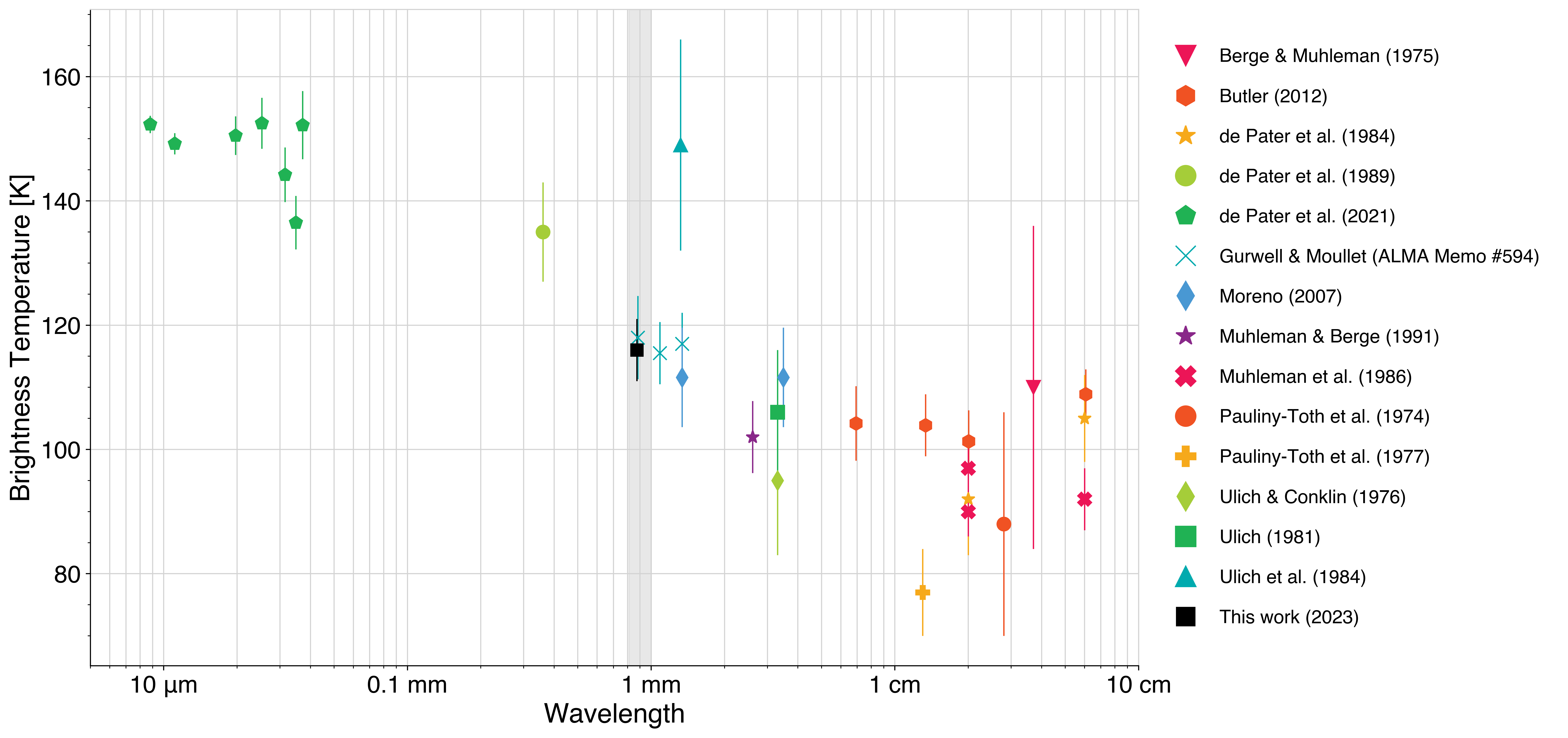}
\caption{Summary of disk-integrated brightness temperature measurements for Callisto plotted as a function of wavelength. The gray vertical bar highlights the wavelength of our measurement, which fits in well with most neighboring data. Data in this plot are taken from: \cite{berge_callisto_1975}, \cite{butler_alma_2012}, \cite{de_pater_vla_1984,de_pater_planetary_1989,de_pater_sofia_2021}, Gurwell \& Moullet (ALMA memo \#594, \citealt{butler_alma_2012}), \cite{moreno_report_2007}, \cite{muhleman_observations_1991}, \cite{muhleman_precise_1986}, \cite{pauliny-toth_brightness_1974,pauliny-toth_observations_1977}, \cite{ulich_observations_1976}, \cite{ulich_millimeter-wavelength_1981}, and  \cite{ulich_planetary_1984}.}
\label{fig:all-disk-integrated}
\end{figure}

We determined the disk-integrated flux density of Callisto’s leading hemisphere is 8.03 $\pm$ 0.40 Jy at 343.5 GHz (0.87 mm), which is equivalent to a brightness temperature of 116 $\pm$ 5 K. In Fig.~\ref{fig:all-disk-integrated}, we place our measurement in context with previous disk-integrated quantities obtained over the mm to cm wavelength regime. Overall, our result agrees well with past work, including the SMA measurements by Gurwell \& Moullet (ALMA memo \#594, \citealt{butler_alma_2012}) collected in the 0.8-1.3 mm range, as well as an IRAM–PdBI observation at 1.3 mm by \cite{moreno_report_2007}. A separate measurement by \cite{ulich_planetary_1984} at 1.3 mm is notably about $\sim$35 K higher than neighboring results. As seen in Fig.~\ref{fig:all-disk-integrated}, there is an overall decrease in brightness temperature with increasing wavelength, with temperatures centered around $\sim$150 K at 10 µm that drop to $\sim$90-100 K longward of 1 cm wavelengths. The trend of brightness temperatures cooling off at longer wavelengths can be attributed to several factors, including the fact that lower frequency observations probe deeper into the subsurface where the diurnal wave is attenuated. Additionally, higher thermal inertia (e.g., denser) materials are often found at depth, and can further bring down daytime brightness temperatures. A similar trend of decreasing brightness temperature with wavelength is observed for the other Galilean satellites (see Fig. 4 of \citealt{de_pater_sofia_2021} and Fig. 6 of \citealt{de_kleer_ganymedes_2021}). 

\begin{figure}
\centering
\includegraphics[scale=0.5]{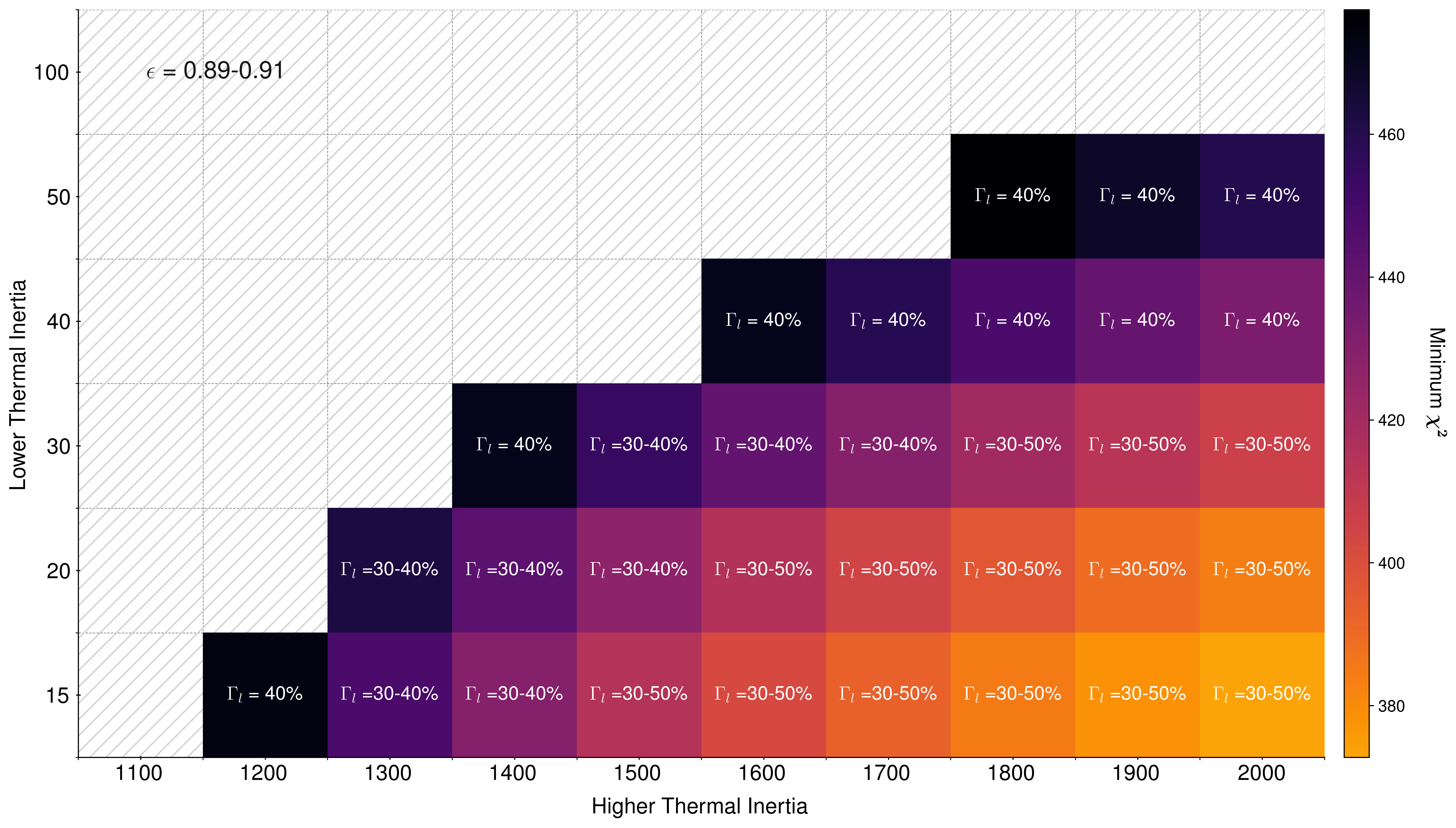}
\caption{ Summary of best-fit two-$\Gamma$ models. Each cell represents the linear mixture results for a combination of $\Gamma_{lower}$ (vertical axis) and $\Gamma_{higher}$ (horizontal axis). The text imposed on each cell gives the range of $\Gamma_{lower}$ (shortened to $\Gamma_{l}$ on the plot) that satisfied the $\chi^2$ cutoff. The actual color of each cell corresponds to the minimum $\chi^2$ achieved within a given mixture range. E.g., the cell corresponding to the mixture of the $\Gamma_{lower}$ = 20 and $\Gamma_{higher}$ = 1600 models is labeled with "$\Gamma_{l}$ = 30-50\%", meaning that mixtures of 30-50\% $\Gamma_{lower}$ = 20 and 70-50\% $\Gamma_{higher}$ = 1600 met the $\chi^2$ constraints. The range of acceptable emissivities is included in the upper left corner of the plot. Empty cells marked with a hatch pattern signal model pairs for which no mixing ratio of thermal inertia components met the $\chi^2$ cutoff. The full range of $\Gamma$ values tested spanned $\Gamma_{lower}$ = 15-400 and $\Gamma_{higher}$ = 500-2000.}
\label{fig:mixture-chi}
\end{figure}

\subsection{Global Properties: Single Thermal Inertia}\label{single-component-results}
We initially fit our ALMA image of Callisto's leading hemisphere using a model parameterized by a single $\Gamma$ and a best-fit emissivity, and found the global properties were poorly constrained using this approach. The single-$\Gamma$ models that best-fit the observed $T_b$ distribution is given by $\Gamma$ = 600-1800, with emissivities of 0.91-0.93. If it is the case that Callisto’s regolith consists of higher-$\Gamma$ material, then constraining precise $\Gamma$ values from the model is difficult because the ALMA data probe materials buried at depths that experience low diurnal wave fluctuations; additionally, increasing $\Gamma$ produces progressively flatter temperature profiles that are difficult to distinguish between.  In addition to varying $\Gamma$, we tested other model parameters, including the porosity, treatment of refraction, and ice to dust ratio; none of these changes provided a significantly better fit. Mainly, the systematic residuals and poor fits, along with the previous evidence \citep{spencer_surfaces_1987} for a thermally heterogeneous surface, motivated our two-component analysis.

\subsection{Global Properties: Two Thermal Inertias}\label{two-component-results}

Given the difficulties of fitting the observed $T_b$ distribution with a single $\Gamma$, we now present the thermal properties acquired by modeling the data with two $\Gamma$ components. The model parameters that satisfied a cutoff of 2$\times$ the minimum chi-squared ($\chi^2$) allow $\Gamma_{lower}$ = 15-50 and $\Gamma_{higher}$ = 1200-2000, with best-fit global spectral emissivities of 0.89-0.91 (Fig.~\ref{fig:mixture-chi}) The acceptable mixing ratios favored the $\Gamma_{higher}$ component, with ranges varying from 70-50\%, compared to 30-50\% for $\Gamma_{lower}$. An example of how the residuals change with varying the percentage of $\Gamma_{lower}$/$\Gamma_{higher}$ is shown in Fig.~\ref{fig:model-comparison}. While the endmember models have $T_b$ deviations from the data up to around $\pm$6 K, the maximum deviations for the best-fitting two-component models are $\sim$2-3 K. Fig.~\ref{fig:model-comparison} also demonstrates how the systematic inability of the single-component models to fit the limbs, especially the morning side, is remediated by the two-$\Gamma$ system. 

Context for Callisto's thermal properties derived at millimeter wavelengths is provided by previous work at infrared wavelengths. Using the IRIS instrument onboard Voyager, \cite{spencer_surfaces_1987} found  Callisto's infrared emission was best-fit by a model that adopted two, vertically stratified thermal inertia components: an upper $\Gamma$ = 15$\substack{+2\\-2}$ layer that controlled the initial fast cooling and a lower $\Gamma$ = 300$\substack{+200\\-200}$ that maintained the slower cooling later during the eclipse. As such, our result that a simple, one-$\Gamma$ model is not sufficient to match Callisto’s thermal millimeter emission is consistent with other work. Similarly, our best-fit $\Gamma_{lower}$ overlaps with the upper layer inferred by \cite{spencer_surfaces_1987}, although there is a larger discrepancy with our inferred second component ($\Gamma_{higher}$ = 1200-2000). However, comparing the thermal properties derived from IRIS data and 345 GHz ALMA data is not straightforward, given that these wavelengths sense the upper mm of a surface and depths of $\sim$5 cm into the subsurface, respectively. Importantly, the thermophysical model treatment by \cite{spencer_surfaces_1987} assumes that the two thermal inertia components are vertically stratified, while ours is implemented with horizontal segregation. \cite{spencer_surfaces_1987} did perform a simple test of horizontally stratified two-$\Gamma$ models, but used $\Gamma$ values that were best-fits for Ganymede and not best-fits for Callisto. Future work using individual ALMA observations at different observing frequencies will provide a more robust evaluation of vertical heterogeneity at millimeter wavelengths.

\begin{figure}
\centering
\includegraphics[scale=0.6]{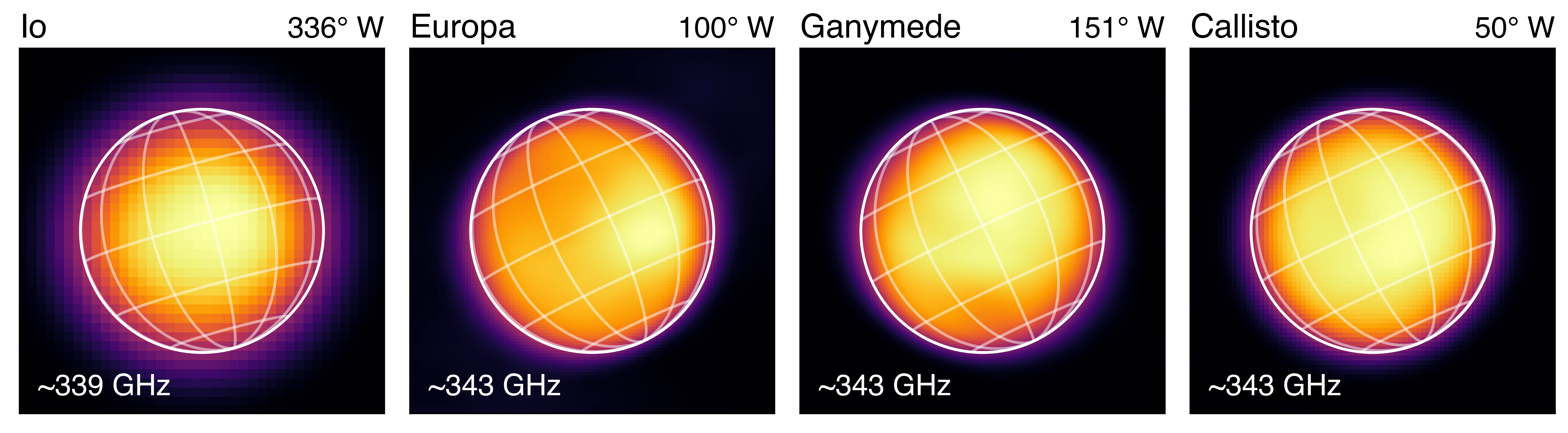}
\caption{Normalized continuum images of the Galilean satellites obtained with ALMA. The longitude indicated on the top right of each plot is the sub-observer longitude at the time of the observation. The latitude/longitude lines are spaced in 30$^{\circ}$ increments. Io data is from \cite{de_pater_alma_2020}, Europa from \cite{thelen_mapping_2023}, Ganymede from \cite{de_kleer_ganymedes_2021}, and Callisto is this work.}
\label{fig:all-sats}
\end{figure}

Additional context for Callisto’s thermal properties derived at millimeter wavelengths is provided by similar ALMA observations of the other Galilean satellites. The mapping of Ganymede by \cite{de_kleer_ganymedes_2021} and Europa by \cite{trumbo_alma_2018} together with the current work on Callisto suggests a general increase of thermal inertia with distance from Jupiter. For example, the $T_b$ distributions on Ganymede observed by \cite{de_kleer_ganymedes_2021} at 343.5, 223, and 97.4 GHz yielded effective thermal inertias of $\Gamma$ = 450$\substack{+300\\-250}$, 350$\substack{+350\\-250}$, and 750$\substack{+200\\-350}$ respectively, while \cite{trumbo_alma_2018} obtained a much lower $\Gamma$ = 95 for Europa with possible surface variations of $\Gamma$ = 40-300. Using eclipse cooling data, \cite{de_pater_alma_2020} inferred Io's thermal inertia to be $\Gamma$ = 50 based on the infrared, and around $\Gamma$ = 350 based on mm data. Regarding horizontal segregation of surface materials, \cite{de_kleer_ganymedes_2021} tested such two-$\Gamma$ models to estimate thermal inertia for observations of Ganymede at separate wavelengths, but such models were not preferred to single $\Gamma$ regimes, which differs from the the proposed explanation for Callisto’s $T_b$ distribution. For emissivities, the measured global values for Callisto of 0.89-0.91 are higher than the 0.75 derived for Europa at 223 GHz \citep{trumbo_alma_2018} and 0.75-0.78 measured for Ganymede at 97.5-343.5 GHz \citep{de_kleer_ganymedes_2021}. This result is reasonable given that the emissivity of ice is lower than rock at millimeter wavelengths and Callisto's surface is not as abundantly ice-rich as its Galilean siblings. Moreover, the spatially-resolved ALMA images of the Galilean moons reveal morphological differences suggestive of their variations in thermal properties (Fig.~\ref{fig:all-sats}). Among the icy Galileans, Callisto’s thermal emission is more homogeneous across varying longitudes than that of either Europa or Ganymede, which is consistent with range of inferred thermal inertias inferred for Callisto reaching higher values. 

\subsection{Local Thermal Residuals}\label{local-residuals}

\begin{figure}
\centering
\includegraphics[scale=0.55]{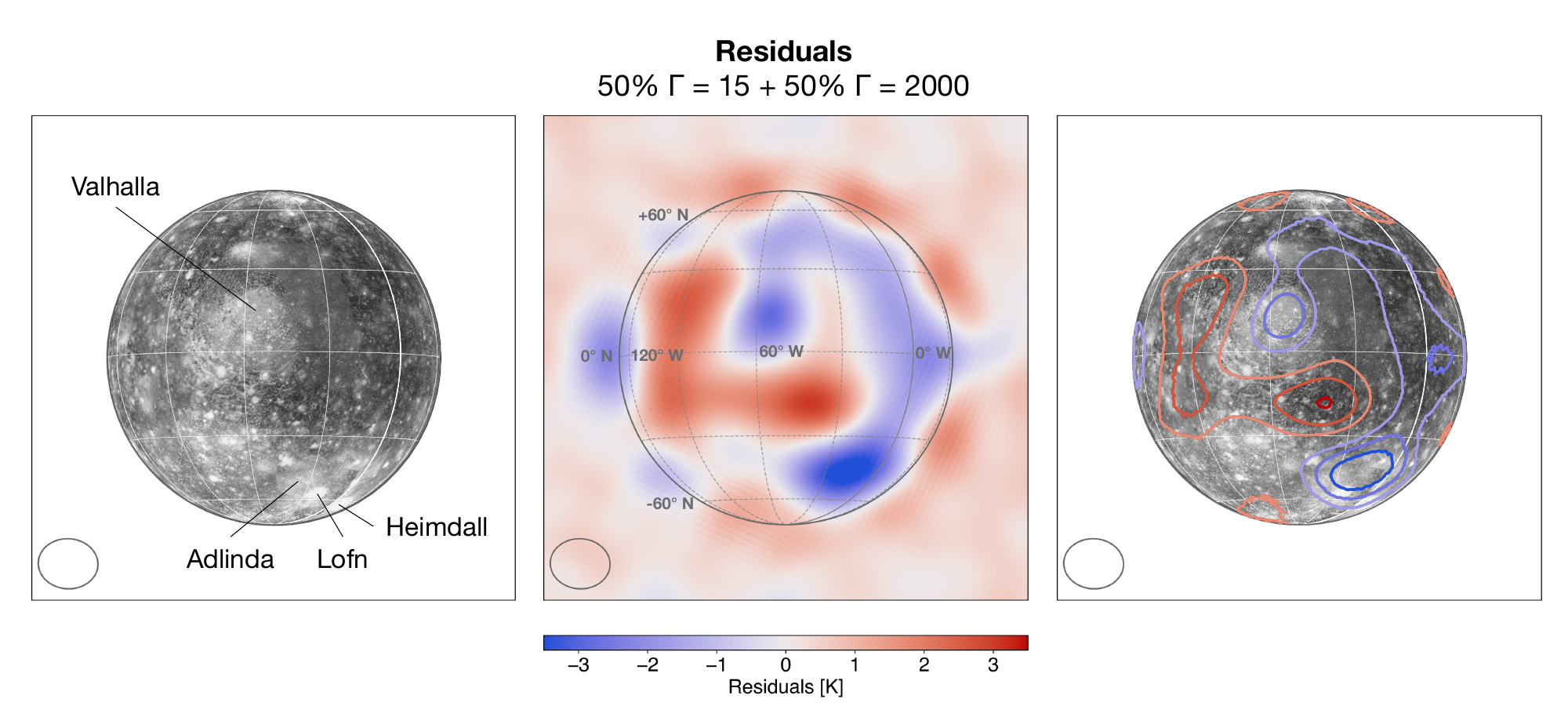}
\caption{A side-by-side comparison of the localized thermal residuals on Callisto with an albedo map projected to the same observer sub-longitude and sub-latitude. The latitude/longitude grid spacing is 30$^{\circ}$. The ellipse in the lower left corner of each panel represents the ALMA synthesized beam at the time of observing. \emph{Left panel}: projected albedo map of Callisto with major geologic features identified. Contrast added for ease of identification. \emph{Middle panel}: Representative residuals from the range of best-fit two thermal inertia models. \emph{Right panel}: Residual contour lines drawn on the projected albedo map to better illustrate the correlation between certain warm/cold regions and local terrain. Contours are drawn for $\pm$ 3, $\pm$ 2, and $\pm$ 1 K.}
\label{fig:localized-residuals}
\end{figure}

In addition to deriving global surface properties from our thermophysical model, we also highlight terrain-specific thermal anomalies that are robust to model parameter changes (Fig.~\ref{fig:localized-residuals}). Note: references to west/east when describing the relative location or orientation of thermal residuals refers to geographic west/east on Callisto, not astronomical west/east.

 \subsubsection{Valhalla Impact Basin}
A noticeably cool region observed in both the calibrated image (Fig.~\ref{fig:data}) and the model residuals (Fig.~\ref{fig:localized-residuals}) is well centered on the Valhalla impact basin. This thermal residual is colder by $\sim$8 K than surrounding terrain in the calibrated image, and about by $\sim$2-3 K compared to model predictions (which incorporate albedo), meaning the SNR of this residual is  $\sim$10 with the 0.2 K measured image rms. Given that the spatial footprint of our ALMA beam at the distance of Callisto is $\sim$790 km, this thermal residual (which is larger than the beam) covers the entirety of Valhalla’s high-albedo $D\sim$ 360 km central zone and part of the $D\sim$ 1900 km inner ridge and trough zone \citep{moore_callisto_2004}. The strength of the cold residual tapers off in outer trough terrain of Valhalla, which extends to $D\sim$ 3800 km \citep{moore_callisto_2004}. In terms of morphology, this thermal feature may be slightly extended to the southwest, which is a  different direction than the ALMA beam elongation and therefore not an artifact of the non-circular resolution element.
	 
A correlation between cold thermal features and impact craters is a phenomenon observed on both Jovian and Saturnian icy satellites. Often, impacts excavate higher thermal inertia materials from the subsurface (e.g., ice-rich, higher density), creating localized terrain that maintains cooler daytime temperatures relative to surface materials that are more responsive to diurnally varying insolation. This is consistent with the thermophysical model residuals at the location of Valhalla improving as higher global thermal inertias were tested, as demonstrated in Fig.~\ref{fig:model-comparison}. Given Valhalla is one of the largest impact features in the Solar System, it is expected that this geologic unit contains relatively more exposed subsurface materials. Impact features on other Galilean moons observed to be thermally-cold via disk-resolved observations include Tros and Osiris on Ganymede \citep{de_kleer_ganymedes_2021,brown_sub-surface_2022}, and Pwyll on Europa \citep{trumbo_alma_2017,trumbo_alma_2018}. Anomalously cold craters have also been observed in the Saturnian system on Titan \citep{janssen_titans_2016} and Rhea \citep{bonnefoy_rheas_2020}. With the results for Valhalla presented here, this general trend is confirmed to extend to the largest class of impacts in the Solar System. 

\subsubsection{Adlinda, Heimdall, and Lofn}
Valhalla is not the only impact feature in our observation co-located with regional thermal anomalies – a collection of three large craters in the southern hemisphere may be the source of the cold residual located near 40$^{\circ}$ S and 12$^{\circ}$ W. This crater complex includes Adlinda and Heimdall, both of which are multi-ring impact features, and Lofn, which sits in-between/on top of the former two. Adlinda is marked by outer troughs that reach up to $D\sim$ 850 km, however almost 1/3 of its structure is obscured by ejecta from Lofn \citep{moore_callisto_2004}. Lofn itself is an exceptionally bright, young anomalous dome crater and is Callisto’s largest non multi-ring feature ($D\sim$ 355 km) \citep{moore_callisto_2004}. The morphology of this crater (e.g., unusually shallow relief) and the presence of water-rich deposits suggest the original Lofn impactor struck a relatively liquid/slushy region of Callisto’s interior \citep{schenk_thickness_2002,greeley_geology_2001, greeley_galileo_2000}. Heimdall bears outer rings that extend up to $D\sim$ 400 km and center geologic units with albedos high enough to saturate the Galileo images \citep{greeley_galileo_2000}. Because these individual craters are below the resolution of this observation, the measured properties are averages between these craters and the surrounding terrains. However, the region can still be seen to possess distinct thermal properties (Fig.~\ref{fig:localized-residuals}).

As with the Valhalla cold region, the cold spot near the Adlinda/Lofn/Heimdall complex could be explained by impact-excavated higher-$\Gamma$ surface products. The residual itself diverges from model predictions by about 3-4 K (so SNR $\sim$15-20) and is roughly 1.5 beams across and is extended toward the northeast. The high albedos of Lofn and Heimdall, as well as spectroscopic evidence, point toward the presence of increased surface ice relative to the surrounding cratered plains, which could result in locally higher thermal inertia. \cite{ligier_new_2020} interpret their near-infrared maps of Callisto obtained using SINFONI on the VLT to suggest the small grains (25-100 µm) abundance is uniquely enhanced near Lofn and Heimdall (and at high latitudes in general). For an icy surface such as Callisto’s, areas with small-grained ice particles can exhibit ice sintering, which increases grain-grain contacts and thereby increases the effective thermal inertia.

Unlike the Valhalla cold residual, however, this cold residual is not well-centered on the three-crater complex--- rather, the temperature minimum is located roughly 12$^{\circ}$ (or roughly $0.08^{\prime\prime}$) northeast of Lofn’s bright central region. Although this region is closer to Callisto’s limb, where the resolution tapers off due to geometric foreshortening, the fact that the residual is extended toward the northeast, e.g. not in the direction of beam elongation, suggests the offset may be real. Moreover, in the visible albedo map (Fig.~\ref{fig:localized-residuals}), there is similar cratered terrain to the west of this residual that diverges from model predictions on order of the 0.2 K image rms, implying the thermal properties of the region corresponding to the cold residual are unique. A potential relationship to this residual may be found in the Galileo NIMS maps of CO\textsubscript{2} 4.2 µm band depths by \cite{hibbitts_distributions_2000}. The vicinity of Heimdall/Lofn bears locally deep CO\textsubscript{2} band depths, with moderate band depths also extending in a northeastern direction. However, because a link is difficult to confirm we only note the spatial correspondence of these features.

\subsubsection{Additional Localized Thermal Anomalies}
In addition to the thermal residuals on Callisto that appear connected to specific geologic units (e.g, Valhalla and the Adlinda/Heimdall/Lofn complex), there are other residuals that are not correlated to distinct geologic units. The first region is a warm, L-shaped residual that occupies the area west and south of Valhalla. The vertical component of the L is centered at almost 90$^{\circ}$ W and is symmetric across the equator with a divergence from the model of $\sim$2-3 K, while the horizontal portion of the L is similarly warm, and extends eastward to $\sim$25$^{\circ}$ W. The reason for the L shape may be because the warm region is cropped by the cold Valhalla impact to the northeast. The second region is the colder area that occupies most of Callisto’s disk east of Valhalla, starting at $\sim$25$^{\circ}$ W and extending toward the limb. According to the geologic map of Callisto \citep{greeley_galileo_2000}, both regions are classified as the same cratered plains terrain, despite exhibiting both cold and warm residuals. This is unlike the other significant thermal residuals shown in Fig.~\ref{fig:localized-residuals}, where the units (i.e. large impacts) are linked to only one type of thermal signature (i.e., cold).  

Given the morphology and location of these residuals, one speculation is that the ALMA data are sensitive to surface texture alterations induced by the Jovian particle environment. In the Jovian and Saturnian systems, the leading hemispheres of the icy satellites are preferentially bombarded by micrometeorites and neutral particles (such as dust), with the satellites’ equatorial to mid-latitude regions most prominently affected, and their trailing hemispheres are preferentially impacted by charged species. Examples of these phenomena include the hemispheric color asymmetry on Europa \citep{pappalardo_europas_2009} and “Pacman”-shaped thermally-cold regions on Mimas and Tethys detected by Cassini \citep{howett_pacman_2012,howett_high-amplitude_2011}. Given that the warm L-shape on Callisto’s leading hemisphere occupies mid-latitude regions near the satellite’s apex of motion, it is possible the local regolith has been texturized to a less-compact material, perhaps due to micrometeorite bombardment since less-compacted regoliths generally have lower thermal inertias and grain contact areas due to lower bulk densities, resulting in faster daytime warming. The micrometeorite bombardment hypothesis is consistent with explanations for similar $T_b$ distributions on Ganymede’s leading hemisphere detected with ALMA \citep{de_kleer_ganymedes_2021}. Moreover, as summarized by \cite{moore_callisto_2004}, solar phase curves produced by \cite{buratti_ganymede_1991} also suggest Callisto’s leading side is less compact than on the trailing side, although \cite{domingue_re-analysis_1997} using opposition effect data from \cite{thompson_photoelectric_1992} infer the opposite result. Although beyond the scope of the current work, a trailing hemisphere image of Callisto at millimeter wavelengths could confirm if the warm region is actually confined to the leading hemisphere. Even if future observations confirm the equatorial warm region is leading hemisphere-specific, it may originate from several, interacting external modification processes.  
\section{Conclusion}\label{Conclusion}
We present a leading hemisphere thermal image of Callisto obtained at millimeter wavelengths (0.87 mm/343 GHz). Our first approach at modeling the thermal emission of Callisto used a thermal model parameterized by a single thermal inertia, for which we obtained a global best-fit of $\Gamma$ = 600-1800 ($\text{J}\:\text{ m}^{-2}\:\text{ K}^{-1}\:\text{ s}^{-1/2}$) and best-fit emissivities of 0.91-0.93. The single-$\Gamma$ approach yielded high and systematic residuals that suggested a two-$\Gamma$ model may describe Callisto’s surface more accurately. Our global best-fit two-$\Gamma$ model adopted $\Gamma_{lower}$ = 15-50 and $\Gamma_{higher}$= 1200-2000, with acceptable mixing ratios of $\Gamma_{lower}$ = 30-50\% and $\Gamma_{higher}$ = 70-50\%, and  emissivities of 0.89-0.91. Our result that a single-$\Gamma$ model is not sufficient to accurately model Callisto’s surface is consistent with results from \cite{spencer_surfaces_1987} derived using IRIS infrared data. The $\Gamma_{higher}$ we derived is higher than properties inferred from the infrared, and the presence of a low-thermal inertia component that may suggest the presence of a very uncompact surface component is consistent with other optical results (e.g., \citealt{buratti_ganymede_1991}). Callisto's thermal inertia components as inferred from these ALMA data are higher than those on Ganymede ($\Gamma \sim$ 450-750; \citealt{de_kleer_ganymedes_2021}) and Europa ($\Gamma \sim$ 95; \citealt{trumbo_alma_2017,trumbo_alma_2018}), which agrees with its more thermally-uniform appearance (Fig.~\ref{fig:all-sats}). 

In addition to deriving a global best-fit to the $T_b$ distribution, we also examined localized residuals to determine correlations, if any, with regional geologic features. We found that the Valhalla impact basin was co-located with a cold temperature anomaly $\sim$3 K, which extends the association between large impact craters/colder temperatures previously observed on other moons up to the multi-ring impact class. In addition, we found that the other primary cold region in our $T_b$ map is in the vicinity of a collection of craters including Adlinda, Heimdall, and Lofn, of which the latter two are among the brightest and youngest large features on the leading hemisphere. However, the cold residual is notably extended away from this suite of craters toward the northeast, which may imply surface textures related to factors other than cratering may be relevant. One residual for which there is no obvious association with a geologic unit is a warm L-shaped region that is nearly centered on the leading hemisphere and is constrained to about $\pm$ 30$^{\circ}$ latitude. We speculated that this feature, like similar warm residuals observed on Ganymede’s leading hemisphere \citep{de_kleer_ganymedes_2021}, might represent a region of lower thermal inertia resulting from micrometeorite bombardment that preferentially texturizes the leading hemisphere. However, such a hypothesis is difficult to confirm in the context of a single thermal observation.

The data presented here provide the first spatially-resolved image of Callisto’s surface at ALMA observing frequencies. Future analysis of observations that include imaging of Callisto’s trailing hemisphere at 0.87 mm/343 GHz, as well as mapping at other thermally-relevant wavelengths (e.g., 100-343 GHz), will help build the global and vertical-depth coverage needed to better understand the thermal features presented here. Additionally, this work offers ground-based observational context for the future European Space Agency (ESA) JUICE (JUpiter Icy Moons Explorer) mission to the Jovian satellites. The JUICE science program highlights Callisto as a key element in crafting the narrative of Jovian satellite formation and the emergence of potentially habitable environments and ocean-bearing worlds \citep{esa_juice_2014}. ALMA imaging of Callisto in the $\sim$100-343 GHz range will provide a near-subsurface ($\sim$cm-m depths) complement to the JUICE instrument payload, which includes ground-penetrating radar (RIME, $\sim$9 MHz, depths of $\sim$9 km) and the Sub-millimeter Wave Instrument (SWI, $\sim$530-601 GHz).

\section*{Acknowledgements}
We acknowledge support from the National Science Foundation Graduate Research Fellowship under Grant No. DGE‐1745301 to M.C., and from the Heising-Simons Foundation via a 51 Pegasi b postdoctoral fellowship and under grant \#2019-1611 to K.dK. Contributions by A.A. were carried out at the Jet Propulsion Laboratory, California Institute of Technology, under a contract with the National Aeronautics and Space Administration (80NM0018D0004). This paper makes use of the following ALMA data: ADS/ JAO.ALMA\#2016.1.00691.S. ALMA is a partnership of ESO (representing its member states), NSF (USA) and NINS (Japan), together with NRC (Canada), MOST and ASIAA (Taiwan), and KASI (Republic of Korea), in cooperation with the Republic of Chile. The Joint ALMA Observatory is operated by ESO, AUI/NRAO, and NAOJ. The National Radio Astronomy Observatory is a facility of the National Science Foundation operated under cooperative agreement by Associated Universities, Inc.
\bibliography{callisto-2023.bib}

\begin{thebibliography}{}
\expandafter\ifx\csname natexlab\endcsname\relax\def\natexlab#1{#1}\fi
\providecommand{\url}[1]{\href{#1}{#1}}
\providecommand{\dodoi}[1]{doi:~\href{http://doi.org/#1}{\nolinkurl{#1}}}
\providecommand{\doeprint}[1]{\href{http://ascl.net/#1}{\nolinkurl{http://ascl.net/#1}}}
\providecommand{\doarXiv}[1]{\href{https://arxiv.org/abs/#1}{\nolinkurl{https://arxiv.org/abs/#1}}}

\bibitem[{Berge \& Muhleman(1975)}]{berge_callisto_1975}
Berge, G.~L., \& Muhleman, D.~O. 1975, Science, 187, 441,
  \dodoi{10.1126/science.187.4175.441}

\bibitem[{Bonnefoy {et~al.}(2020)Bonnefoy, Le~Gall, Lellouch, Leyrat, Janssen,
  \& Sultana}]{bonnefoy_rheas_2020}
Bonnefoy, L., Le~Gall, A., Lellouch, E., {et~al.} 2020, Icarus, 352, 113947,
  \dodoi{10.1016/j.icarus.2020.113947}

\bibitem[{Bottke {et~al.}(2013)Bottke, Vokrouhlický, Nesvorný, \&
  Moore}]{bottke_black_2013}
Bottke, W.~F., Vokrouhlický, D., Nesvorný, D., \& Moore, J.~M. 2013, Icarus,
  223, 775, \dodoi{10.1016/j.icarus.2013.01.008}

\bibitem[{Brogan {et~al.}(2018)Brogan, Hunter, \&
  Fomalont}]{brogan_advanced_2018}
Brogan, C.~L., Hunter, T.~R., \& Fomalont, E.~B. 2018, arXiv:1805.05266
  [astro-ph].
\newblock \url{http://arxiv.org/abs/1805.05266}

\bibitem[{Brown {et~al.}(2022)Brown, Bolton, Misra, Levin, Zhang, Lunine,
  Stevenson, \& Siegler}]{brown_sub-surface_2022}
Brown, S., Bolton, S., Misra, S., {et~al.} 2022, Sub-surface {Observations} of
  {Ganymede}'s {Ice} {Shell} from the {Juno} {Microwave} {Radiometer}, Tech.
  Rep. EGU22-10748, Copernicus Meetings, \dodoi{10.5194/egusphere-egu22-10748}

\bibitem[{Buratti(1991)}]{buratti_ganymede_1991}
Buratti, B.~J. 1991, Icarus, 92, 312, \dodoi{10.1016/0019-1035(91)90054-W}

\bibitem[{Butler(2012)}]{butler_alma_2012}
Butler, B. 2012, {ALMA} {Memo} 594, Tech. rep.
\newblock \url{https://science.nrao.edu/facilities/alma/
  aboutALMA/Technology/ALMA_Memo_Series/alma594/abs594}

\bibitem[{Butler \& Bastian(1999)}]{taylor_solar_1999}
Butler, B.~J., \& Bastian, T. 1999, in Synthesis {Imaging} in {Radio}
  {Astronomy} {II}, ed. G.~B. Taylor, C.~L. Carilli, \& R.~A. Perley, Vol. 180.
\newblock \url{https://ui.adsabs.harvard.edu/abs/1999ASPC..180.....T}

\bibitem[{Carlson {et~al.}(2009)Carlson, Calvin, Dalton, Hansen, Hudson,
  Johnson, McCord, \& Moore}]{pappalardo_europas_2009}
Carlson, R.~W., Calvin, W.~M., Dalton, J.~B., {et~al.} 2009, in Europa, ed.
  R.~T. Pappalardo, W.~B. McKinnon, \& K.~K. Khurana (University of Arizona
  Press), \dodoi{10.2307/j.ctt1xp3wdw}

\bibitem[{Clark(1980)}]{clark_efficient_1980}
Clark, B.~G. 1980, Astronomy and Astrophysics, 89, 377.
\newblock \url{https://ui.adsabs.harvard.edu/abs/1980A&A....89..377C/abstract}

\bibitem[{Clark \& McCord(1980)}]{clark_galilean_1980}
Clark, R.~N., \& McCord, T.~B. 1980, Icarus, 41, 323

\bibitem[{Davies {et~al.}(2006)Davies, Wilson, Matson, Leone, Keszthelyi, \&
  Jaeger}]{davies_heartbeat_2006}
Davies, A.~G., Wilson, L., Matson, D., {et~al.} 2006, Icarus, 184, 460,
  \dodoi{10.1016/j.icarus.2006.05.012}

\bibitem[{de~Kleer {et~al.}(2021{\natexlab{a}})de~Kleer, Butler, de~Pater,
  Gurwell, Moullet, Trumbo, \& Spencer}]{de_kleer_ganymedes_2021}
de~Kleer, K., Butler, B., de~Pater, I., {et~al.} 2021{\natexlab{a}}, The
  Planetary Science Journal, 2, 5, \dodoi{10.3847/PSJ/abcbf4}

\bibitem[{de~Kleer {et~al.}(2021{\natexlab{b}})de~Kleer, Cambioni, \&
  Shepard}]{de_kleer_surface_2021}
de~Kleer, K., Cambioni, S., \& Shepard, M. 2021{\natexlab{b}}, The Planetary
  Science Journal, 2, 149, \dodoi{10.3847/PSJ/ac01ec}

\bibitem[{de~Kleer {et~al.}(2019)de~Kleer, de~Pater, Molter, Banks, Davies,
  Alvarez, Campbell, Aycock, Pelletier, Stickel, Kacprzak, Nielsen, Stern, \&
  Tollefson}]{de_kleer_ios_2019}
de~Kleer, K., de~Pater, I., Molter, E.~M., {et~al.} 2019, The Astronomical
  Journal, 158, 29, \dodoi{10.3847/1538-3881/ab2380}

\bibitem[{de~Pater {et~al.}(1984)de~Pater, Brown, \&
  Dickel}]{de_pater_vla_1984}
de~Pater, I., Brown, R.~A., \& Dickel, J.~R. 1984, Icarus, 57, 93,
  \dodoi{10.1016/0019-1035(84)90011-3}

\bibitem[{de~Pater {et~al.}(2021)de~Pater, Fletcher, Reach, Goullaud, Orton,
  Wong, \& Gehrz}]{de_pater_sofia_2021}
de~Pater, I., Fletcher, L.~N., Reach, W.~T., {et~al.} 2021, The Planetary
  Science Journal, 2, 226, \dodoi{10.3847/PSJ/ac2d24}

\bibitem[{de~Pater {et~al.}(2020)de~Pater, Luszcz-Cook, Rojo, Redwing, Kleer,
  \& Moullet}]{de_pater_alma_2020}
de~Pater, I., Luszcz-Cook, S., Rojo, P., {et~al.} 2020, The Planetary Science
  Journal, 1, 60, \dodoi{10.3847/PSJ/abb93d}

\bibitem[{de~Pater {et~al.}(1989)de~Pater, Ulich, Kreysa, \&
  Chini}]{de_pater_planetary_1989}
de~Pater, I., Ulich, B.~L., Kreysa, E., \& Chini, R. 1989, Icarus, 79, 190

\bibitem[{Domingue \& Verbiscer(1997)}]{domingue_re-analysis_1997}
Domingue, D., \& Verbiscer, A. 1997, Icarus, 128, 49,
  \dodoi{10.1006/icar.1997.5730}

\bibitem[{ESA(2014)}]{esa_juice_2014}
ESA. 2014, {JUICE} {Definition} {Study} {Report}, Tech. rep.

\bibitem[{Ferrari \& Lucas(2016)}]{ferrari_low_2016}
Ferrari, C., \& Lucas, A. 2016, Astronomy \& Astrophysics, 588, A133,
  \dodoi{10.1051/0004-6361/201527625}

\bibitem[{Greeley {et~al.}(2004)Greeley, Chyba, Head, McCord, McKinnon,
  Pappalardo, \& Figueredo}]{bagenal_geology_2004}
Greeley, R., Chyba, C., Head, J.~W., {et~al.} 2004, in Jupiter: {The} {Planet},
  {Satellites} and {Magnetosphere}, ed. F.~Bagenal, {Timothy Dowling}, \& W.~B.
  McKinnon (Cambridge University Press)

\bibitem[{Greeley {et~al.}(2001)Greeley, Heiner, \&
  Klemaszewski}]{greeley_geology_2001}
Greeley, R., Heiner, S., \& Klemaszewski, J.~E. 2001, Journal of Geophysical
  Research: Planets, 106, 3261, \dodoi{10.1029/2000JE001262}

\bibitem[{Greeley {et~al.}(2000)Greeley, Klemaszewski, \&
  Wagner}]{greeley_galileo_2000}
Greeley, R., Klemaszewski, J., \& Wagner, R. 2000, Planetary and Space Science,
  48, 829, \dodoi{10.1016/S0032-0633(00)00050-7}

\bibitem[{Grundy {et~al.}(1999)Grundy, Buie, Stansberry, Spencer, \&
  Schmitt}]{grundy_near-infrared_1999}
Grundy, W.~M., Buie, M.~W., Stansberry, J.~A., Spencer, J.~R., \& Schmitt, B.
  1999, Icarus, 142, 536, \dodoi{10.1006/icar.1999.6216}

\bibitem[{Hayne {et~al.}(2017)Hayne, Bandfield, Siegler, Vasavada, Ghent,
  Williams, Greenhagen, Aharonson, Elder, Lucey, \& Paige}]{hayne_global_2017}
Hayne, P.~O., Bandfield, J.~L., Siegler, M.~A., {et~al.} 2017, Journal of
  Geophysical Research: Planets, 122, 2371, \dodoi{10.1002/2017JE005387}

\bibitem[{Hibbitts {et~al.}(2000)Hibbitts, McCord, \&
  Hansen}]{hibbitts_distributions_2000}
Hibbitts, C.~A., McCord, T.~B., \& Hansen, G.~B. 2000, Journal of Geophysical
  Research: Planets, 105, 22541, \dodoi{10.1029/1999JE001101}

\bibitem[{Howett {et~al.}(2012)Howett, Spencer, Hurford, Verbiscer, \&
  Segura}]{howett_pacman_2012}
Howett, C., Spencer, J., Hurford, T., Verbiscer, A., \& Segura, M. 2012,
  Icarus, 221, 1084, \dodoi{10.1016/j.icarus.2012.10.013}

\bibitem[{Howett {et~al.}(2019)Howett, Spencer, Hurford, Verbiscer, \&
  Segura}]{howett_maps_2019}
---. 2019, Icarus, 321, 705, \dodoi{10.1016/j.icarus.2018.12.018}

\bibitem[{Howett {et~al.}(2011)Howett, Spencer, Schenk, Johnson, Paranicas,
  Hurford, Verbiscer, \& Segura}]{howett_high-amplitude_2011}
Howett, C., Spencer, J., Schenk, P., {et~al.} 2011, Icarus, 216, 221,
  \dodoi{10.1016/j.icarus.2011.09.007}

\bibitem[{Janssen {et~al.}(2016)Janssen, Le~Gall, Lopes, Lorenz, Malaska,
  Hayes, Neish, Solomonidou, Mitchell, Radebaugh, Keihm, Choukroun, Leyrat,
  Encrenaz, \& Mastrogiuseppe}]{janssen_titans_2016}
Janssen, M., Le~Gall, A., Lopes, R., {et~al.} 2016, Icarus, 270, 443,
  \dodoi{10.1016/j.icarus.2015.09.027}

\bibitem[{Johnson {et~al.}(1983)Johnson, Soderblom, Mosher, Danielson, Cook, \&
  Kupferman}]{johnson_global_1983}
Johnson, T.~V., Soderblom, L.~A., Mosher, J.~A., {et~al.} 1983, Journal of
  Geophysical Research: Solid Earth, 88, 5789, \dodoi{10.1029/JB088iB07p05789}

\bibitem[{Ligier {et~al.}(2020)Ligier, Calvin, Carter, Poulet, Paranicas, \&
  Snodgrass}]{ligier_new_2020}
Ligier, N., Calvin, W.~M., Carter, J., {et~al.} 2020, LPSC Contrib. No. 1959

\bibitem[{McMullin {et~al.}(2007)McMullin, Waters, Schiebel, Young, \&
  Golap}]{mcmullin_casa_2007}
McMullin, J.~P., Waters, B., Schiebel, D., Young, W., \& Golap, K. 2007, 376,
  127.
\newblock \url{https://ui.adsabs.harvard.edu/abs/2007ASPC..376..127M}

\bibitem[{Moore {et~al.}(2004)Moore, Chapman, Bierhaus, Greeley, Chuang,
  Klemaszewski, Clark, Dalton, Hibbitts, Schenk, Spencer, \&
  Wagner}]{moore_callisto_2004}
Moore, J., Chapman, C.~R., Bierhaus, E.~B., {et~al.} 2004, in Jupiter: {The}
  {Planet}, {Satellites} and {Magnetosphere} (Cambridge: Cambridge University
  Press), 397--427

\bibitem[{Moreno(2007)}]{moreno_report_2007}
Moreno, R. 2007, Report on continuum measurements of {Ganymede} and {Callisto}
  with the {IRAM}–{PdB} interferometer : {Application} to flux calibration,
  {Internal} {Memo}, Tech. rep.

\bibitem[{Morrison {et~al.}(1972)Morrison, Cruikshank, \&
  Murphy}]{morrison_temperatures_1972}
Morrison, D., Cruikshank, D.~P., \& Murphy, R.~E. 1972, The Astrophysical
  Journal, 173, L143, \dodoi{10.1086/180934}

\bibitem[{Muders {et~al.}(2014)Muders, Wyrowski, Lightfoot, Williams, Nakazato,
  Kosugi, Davis, \& Kern}]{muders_alma_2014}
Muders, D., Wyrowski, F., Lightfoot, J., {et~al.} 2014, in Astronomical {Data}
  {Analysis} {Software} and {Systems} {XXIII}, ed. N.~Manset \& P.~Forshay,
  Vol. 485, San Francisco, CA, 383

\bibitem[{Muhleman \& Berge(1991)}]{muhleman_observations_1991}
Muhleman, D.~O., \& Berge, G.~L. 1991, Icarus, 92, 263,
  \dodoi{10.1016/0019-1035(91)90050-4}

\bibitem[{Muhleman {et~al.}(1986)Muhleman, Berge, Rudy, \&
  Niell}]{muhleman_precise_1986}
Muhleman, D.~O., Berge, G.~L., Rudy, D., \& Niell, A.~E. 1986, The Astronomical
  Journal, 92, 1428, \dodoi{10.1086/114279}

\bibitem[{Mura {et~al.}(2020)Mura, Adriani, Tosi, Lopes, Sindoni, Filacchione,
  Williams, Davies, Plainaki, Bolton, Altieri, Cicchetti, Grassi, Migliorini,
  Moriconi, Noschese, Olivieri, Piccioni, \& Sordini}]{mura_infrared_2020}
Mura, A., Adriani, A., Tosi, F., {et~al.} 2020, Icarus, 341, 113607,
  \dodoi{10.1016/j.icarus.2019.113607}

\bibitem[{Pappalardo {et~al.}(2004)Pappalardo, Collins, Head, Helfenstein,
  McCord, Moore, Prockter, Schenk, \& Spencer}]{bagenal_ganymede_2004}
Pappalardo, R., Collins, G., Head, J.~W., {et~al.} 2004, in Jupiter: {The}
  {Planet}, {Satellites} and {Magnetosphere}, ed. F.~Bagenal, {Timothy
  Dowling}, \& W.~B. McKinnon (Cambridge University Press)

\bibitem[{Pauliny-Toth {et~al.}(1974)Pauliny-Toth, Witzel, \&
  Gorgolewski}]{pauliny-toth_brightness_1974}
Pauliny-Toth, I. I.~K., Witzel, A., \& Gorgolewski, S. 1974, Astronomy and
  Astrophysics, Vol. 34, p. 129 (1974), 34, 129.
\newblock
  \url{https://ui.adsabs.harvard.edu/abs/1974A%26A....34..129P/abstract}

\bibitem[{Pauliny-Toth {et~al.}(1977)Pauliny-Toth, Witzel, \&
  Gorgolewski}]{pauliny-toth_observations_1977}
---. 1977, Astronomy and Astrophysics, 58, L27.
\newblock \url{https://ui.adsabs.harvard.edu/abs/1977A&A....58L..27P/abstract}

\bibitem[{Rau \& Cornwell(2011)}]{rau_multi-scale_2011}
Rau, U., \& Cornwell, T.~J. 2011, Astronomy \& Astrophysics, 532, A71,
  \dodoi{10.1051/0004-6361/201117104}

\bibitem[{Sault \& Wieringa(1994)}]{sault_multi-frequency_1994}
Sault, R.~J., \& Wieringa, M.~H. 1994, Astronomy and Astrophysics Supplement
  Series, 108, 585.
\newblock \url{https://ui.adsabs.harvard.edu/abs/1994A&AS..108..585S}

\bibitem[{Schaible {et~al.}(2017)Schaible, Johnson, Zhigilei, \&
  Piqueux}]{schaible_high_2017}
Schaible, M.~J., Johnson, R.~E., Zhigilei, L.~V., \& Piqueux, S. 2017, Icarus,
  285, 211, \dodoi{10.1016/j.icarus.2016.08.033}

\bibitem[{Schenk {et~al.}(2011)Schenk, Hamilton, Johnson, McKinnon, Paranicas,
  Schmidt, \& Showalter}]{schenk_plasma_2011}
Schenk, P., Hamilton, D.~P., Johnson, R.~E., {et~al.} 2011, Icarus, Volume 211,
  Issue 1, p. 740-757., 211, 740, \dodoi{10.1016/j.icarus.2010.08.016}

\bibitem[{Schenk(1995)}]{schenk_geology_1995}
Schenk, P.~M. 1995, Journal of Geophysical Research: Planets, 100, 19023,
  \dodoi{10.1029/95JE01855}

\bibitem[{Schenk(2002)}]{schenk_thickness_2002}
---. 2002, Nature, 417, 419, \dodoi{10.1038/417419a}

\bibitem[{Spencer(1987)}]{spencer_surfaces_1987}
Spencer, J.~R. 1987, PhD thesis, University of Arizona.
\newblock \url{https://repository.arizona.edu/handle/10150/184098}

\bibitem[{Thelen {et~al.}(2023)Thelen, de~Kleer, Camarca, Butler, de~Pater,
  Gurwell, \& Moullet}]{thelen_mapping_2023}
Thelen, A., de~Kleer, K., Camarca, M., {et~al.} 2023, LPSC Contrib. No. 2806

\bibitem[{Thompson \& Lockwood(1992)}]{thompson_photoelectric_1992}
Thompson, D.~T., \& Lockwood, G.~W. 1992, Journal of Geophysical Research, 97,
  14761, \dodoi{10.1029/92JE01399}

\bibitem[{Trumbo {et~al.}(2017)Trumbo, Brown, \& Butler}]{trumbo_alma_2017}
Trumbo, S.~K., Brown, M.~E., \& Butler, B.~J. 2017, The Astronomical Journal,
  154, 148, \dodoi{10.3847/1538-3881/aa8769}

\bibitem[{Trumbo {et~al.}(2018)Trumbo, Brown, \& Butler}]{trumbo_alma_2018}
---. 2018, The Astronomical Journal, 156, 161, \dodoi{10.3847/1538-3881/aada87}

\bibitem[{Ulich(1981)}]{ulich_millimeter-wavelength_1981}
Ulich, B.~L. 1981, The Astronomical Journal, 86, 1619, \dodoi{10.1086/113046}

\bibitem[{Ulich \& Conklin(1976)}]{ulich_observations_1976}
Ulich, B.~L., \& Conklin, E.~K. 1976, Icarus, 27, 183,
  \dodoi{10.1016/0019-1035(76)90001-4}

\bibitem[{Ulich {et~al.}(1984)Ulich, Duckel, \&
  de~Pater}]{ulich_planetary_1984}
Ulich, B.~L., Duckel, J.~R., \& de~Pater, I. 1984, Icarus, 60, 590,
  \dodoi{10.1016/0019-1035(84)90166-0}

\bibitem[{Zimmer(2000)}]{zimmer_subsurface_2000}
Zimmer, C. 2000, Icarus, 147, 329, \dodoi{10.1006/icar.2000.6456}

\end{thebibliography}
\bibliographystyle{aasjournal}

\end{document}